\documentclass{aa}

\usepackage{psfig, graphics}
\usepackage{txfonts}
\usepackage[dvips]{graphicx}
\usepackage[authoryear]{natbib}
\usepackage[latin1]{inputenc}

\title{Quasi full-disk maps of solar horizontal velocities using SDO/HMI data}

\author{ Th.~Roudier\inst{1,2}, M.~Rieutord\inst{1,2}, J.M.~Malherbe\inst{3}, N.~Renon \inst{4}, 
T.~Berger\inst{5}, Z.~Frank\inst{5},  V.~Prat\inst{1,2}, L.~Gizon\inst{6,7}, M.~\v{S}vanda\inst{6,8,9}}

\date{Received \today  / Submitted }

\offprints{Th. Roudier}
 
\institute{Universit\'e de Toulouse; UPS-OMP; IRAP; Toulouse, France
\and CNRS; IRAP; 14, avenue Edouard Belin, F-31400 Toulouse, France
\and LESIA, Observatoire de Paris, Section de Meudon, 92195 Meudon, France
\and CALMIP, DTSI Universit\'e Paul Sabatier, Universit\'e de Toulouse 31062 Toulouse, France
\and Lockheed Martin Advance Technology Center, Palo Alto, CA, USA
\and Max-Planck-Institut f\"ur Sonnensystemforschung, Max-Planck-Strasse
 2, 37191 Katlenburg-Lindau, Germany. 
\and Institut  f\"ur Astrophysik, Georg-August Universit\"at G\"ottingen, 37077 G\"ottingen, Germany. 
\and Astronomical Institute, Academy of Sciences of the Czech Republic (v.~v.~i.), Fri\v{c}ova 298, 
CZ-25165 Ond\v{r}ejov, Czech Republic
\and Astronomical Institute, Faculty of Mathematics and Physics, Charles University in Prague, 
V Hole\v{s}ovi\v{c}k\'ach 2, CZ-18000 Prague~8, Czech Republic}

\begin{document}

\authorrunning{Roudier et al.}
\titlerunning{Quasi full-disk maps of solar horizontal velocities using SDO/HMI data.}

\abstract{% Context
}{%Aims 
 For the first time, the motion of granules (solar plasma on the surface on scales larger than 2.5 Mm) has been followed over the entire visible 
surface of the Sun, using SDO/HMI white-light data. 
}{%Method
 Horizontal velocity fields are derived from image correlation tracking using a new  
version of the coherent structure tracking algorithm.The spatial and temporal resolutions of the 
horizontal velocity map are 2.5 Mm and 30 min respectively .
}{% Results
 From this reconstruction, using the multi-resolution analysis, one can obtain to the 
velocity field at different scales with its derivatives such as the horizontal divergence or 
the vertical component of the vorticity. 
The intrinsic error on the velocity is $\sim$ 0.25 km~s$^{-1}$ for a time sequence of 30 minutes and 
a mesh size of 2.5 Mm.This is acceptable compared to the granule velocities, which range between  
0.3 km~s$^{-1}$ and 1.8 km~s$^{-1}$. A high correlation between velocities computed from \textit{Hinode} and  
SDO/HMI has been found (85\%).  From the data we derive the power spectrum of the 
supergranulation horizontal velocity field, the solar differential rotation, and  the meridional
velocity.
}{% Conclusions 
}

\keywords{The Sun: Atmosphere -- The Sun: Granulation}
\maketitle

\section{Introduction}

The Sun is a star whose plasma flows are the main source of its
magnetic evolution and activity. The coupling of these motions and the
magnetic field in the convection zone is the driver of the magnetic
activity through the dynamo process. The description of the physical
properties of the convective zone requires the knowledge of these motions
at all scales in space and time.

Determining plasma motions inside the Sun is a difficult
task. While local helioseismology can be used to infer vector flows in
three dimensions just below the solar surface, this is only possible for
spatial scales larger than 5 Mm and temporal scales longer than several
hours \citep{Jackie08,Gizon2010,svanda11}.

From the analysis of the proper motion of photospheric structures  (solar granules), it has 
been shown \citep{RRLNS01} that it is possible to get the horizontal plasma flow on the Sun surface. 
In more detail, using numerical simulation in a large horizontal box, \cite{RRLNS01} show that 
granule motions are highly correlated with horizontal flows when the 
scale is larger than $\sim 2500$~km; at smaller scale, granule motions should be 
considered as (solar) turbulent noise. Such techniques have been used on 
relatively small fields of view (few arcminutes) and usually located at the disk center.

Very recently, \textit{Hinode} and SDO satellites produced 
time sequences of images of the solar surface without  disturbance
from the terrestrial atmosphere (seeing).  These observations allow us  to simultaneously
measure horizontal velocities at different
spatial resolutions and to study long-time sequences of velocities over the full Sun
(HMI/SDO).

In this paper, we describe a method  for determining horizontal velocities from proper 
motions of the solar structures observed on the full-disk Sun. For the first time, the motion of granules (solar
plasma flows on the surface on scales larger than 2.5 Mm) has been followed over the 
full visible Sun surface. Horizontal velocity fields are derived from granule tracking using a new 
version of the coherent structure tracking (CST) algorithm \citep{RRRD07}.

We first present in some detail the algorithm based on granule tracking, 
which is able to give a reconstruction of the velocity field at all
scales larger than the sampling scale. The CST also offers the possibility of
selecting specific structures according to their nature, size, lifetime,
etc. and study their motion. In the next section we discuss the different
steps of the algorithm, the segmentation and interpolation processes, and
a comparison of \textit{Hinode} and SDO flow maps. 
 We illustrate the method with applications 
such as the calculation of the kinetic power spectrum of supergranulation, 
and the measurement of the solar differential rotation and meridional flow 
at the central meridian. Discussion and conclusions follow.

\begin{figure*}
\centerline{\resizebox{9cm}{9cm}{\includegraphics{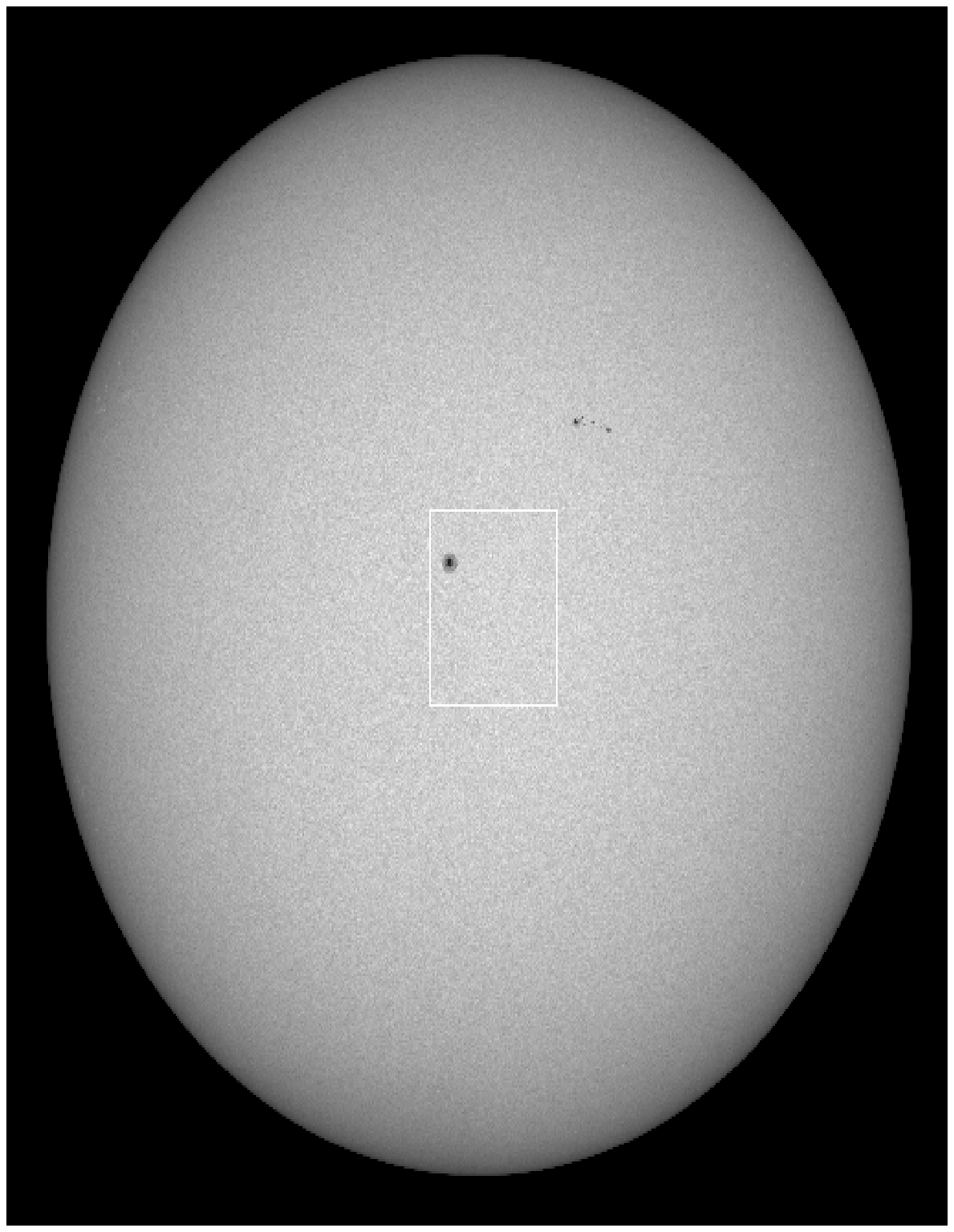}}
\resizebox{9cm}{9cm}{\includegraphics{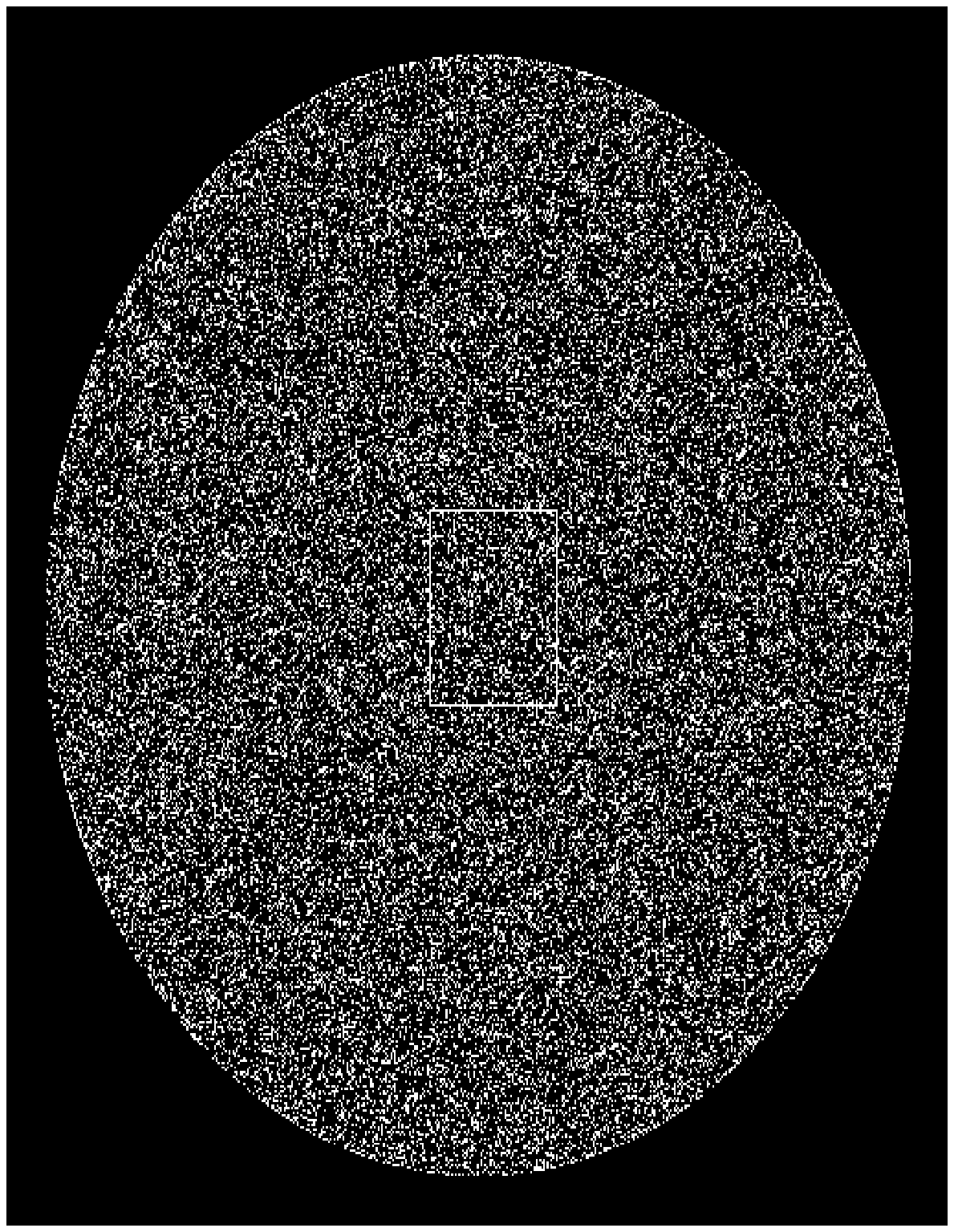}}}
\caption[]{Full Sun HMI/SDO white-light  on August 30, 2010 (left) and
the segmented map, in which about 500 000 granules are detected (right).}
\label{full_sun}
\end{figure*}

\section{Observations}

\subsection{\textit{Hinode} observations}

We used data sets of the Solar Optical Telescope (SOT), on-board the
\textit{Hinode}\footnote{The \textit{Hinode} spacecraft launched in 2006,
was designed and is now operated by JAXA (Japanese Space Authorities) in
cooperation with NASA (National Aeronautics and Space Administration) and
ESA (European Space Agency).} mission \citep[e.g.][]{STISO08,ITSSO04}.
The SOT has a 50~cm primary mirror with a spatial resolution of about 
0.2~\arcsec at 550~nm. For our study, we used blue continuum observations
at 450.45~nm from the \textit{Hinode}/SOT BFI (BroadBand Filter Imager).
The observations were recorded on August 30 2010 from
08:04:33 to 10:59:36 UT.  Solar rotation is compensated 
to observe exactly the same region of the Sun.  The time step is 60 sec
and  the field of view is $76.7~\arcsec\times76.7~\arcsec$ with a pixel scale 
of $0\farcs1089$. After alignment, the useful field-of-view is reduced to
$76\farcs0~\times~74\farcs9$.

To remove the effects of the oscillations, we applied a subsonic
Fourier filter.  This filter is defined by a cone in the $k-\omega$
space, where $k$ and $\omega$ are spatial and temporal frequencies. All
Fourier components with $\omega/k\leq V=7\,\mathrm{km~s}^{-1}$  were retained 
to keep only convective  motions  \citep[][]{TTTFS89}.

\subsection{SDO-HMI observations}

The Helioseismic and Magnetic Imager onboard the Solar Dynamics Observatory
(HMI/SDO) yields uninterrupted high-resolution observations of the
entire disk. This gives a unique opportunity for mapping surface flows on
various scales (spatial and temporal). Using the HMI/SDO white-light data
from  August 30 2010, we derived horizontal velocity fields from image
granulation tracking using a newly developed version of the CST algorithm. 
The time step is 45 seconds with a pixel scale of $0\farcs5$. HMI/SDO white-light 
data from April 10 2010, July 10  2010 and  May 12,13
2011 are used to measure solar differential rotation during 24h for each
of them.

\section{The new CST method}

Granule tracking is a difficult task because of the complex evolution on
granules depending of their size. Granule fragmentation and mixing must be
managed very carefully to avoid the generating noise. We pursue here
the development of a method called coherent structure tracking, or CST,
which determines the horizontal motion of granules in the field of view
\citep{RRRD07}. The preceding version was developed for a field of a 
few arcminutes, generally located at the disk center of the Sun  where the
solar rotation was removed by alignment of the frames.
 The application of the CST to HMI/SDO data requires a new granule time
 labeling methode which takes care of the motion caused by the solar rotation to avoid 
misidentification.  Indeed, in the preceding CST version the time labeling 
was processed by following the barycenter of the granule. The solar rotation that
is present in HMI/SDO data and the evolution of granules like the birth of 
a new (small) granule close to an existing granule between two frames can lead to
misidentification of barycenters. This leads to a bad temporal labeling and 
generates noticeable noise in the final velocity maps.

\begin{figure*}
\resizebox{9.34cm}{9.3cm}{\includegraphics{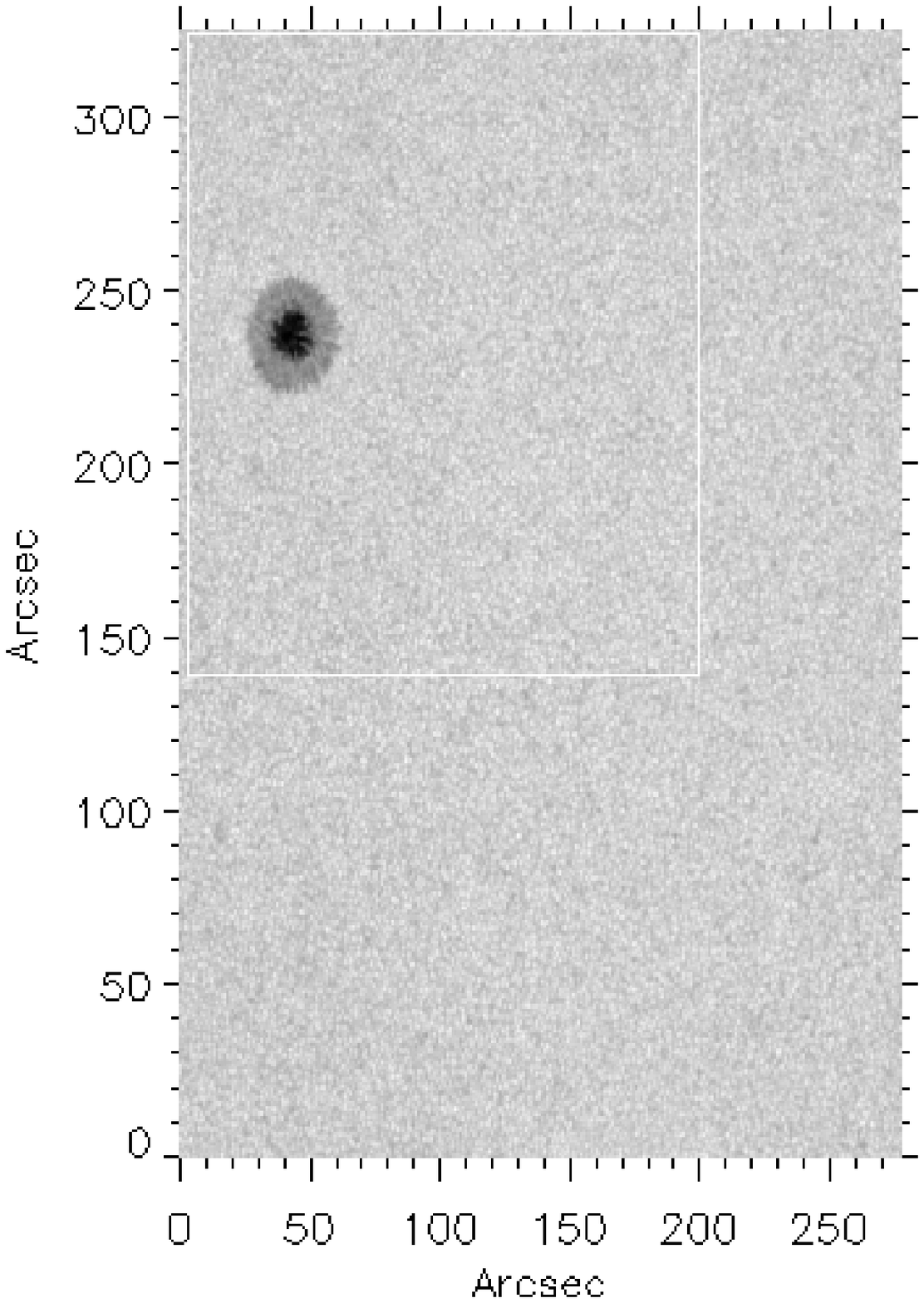}}
\resizebox{9.34cm}{9.3cm}{\includegraphics{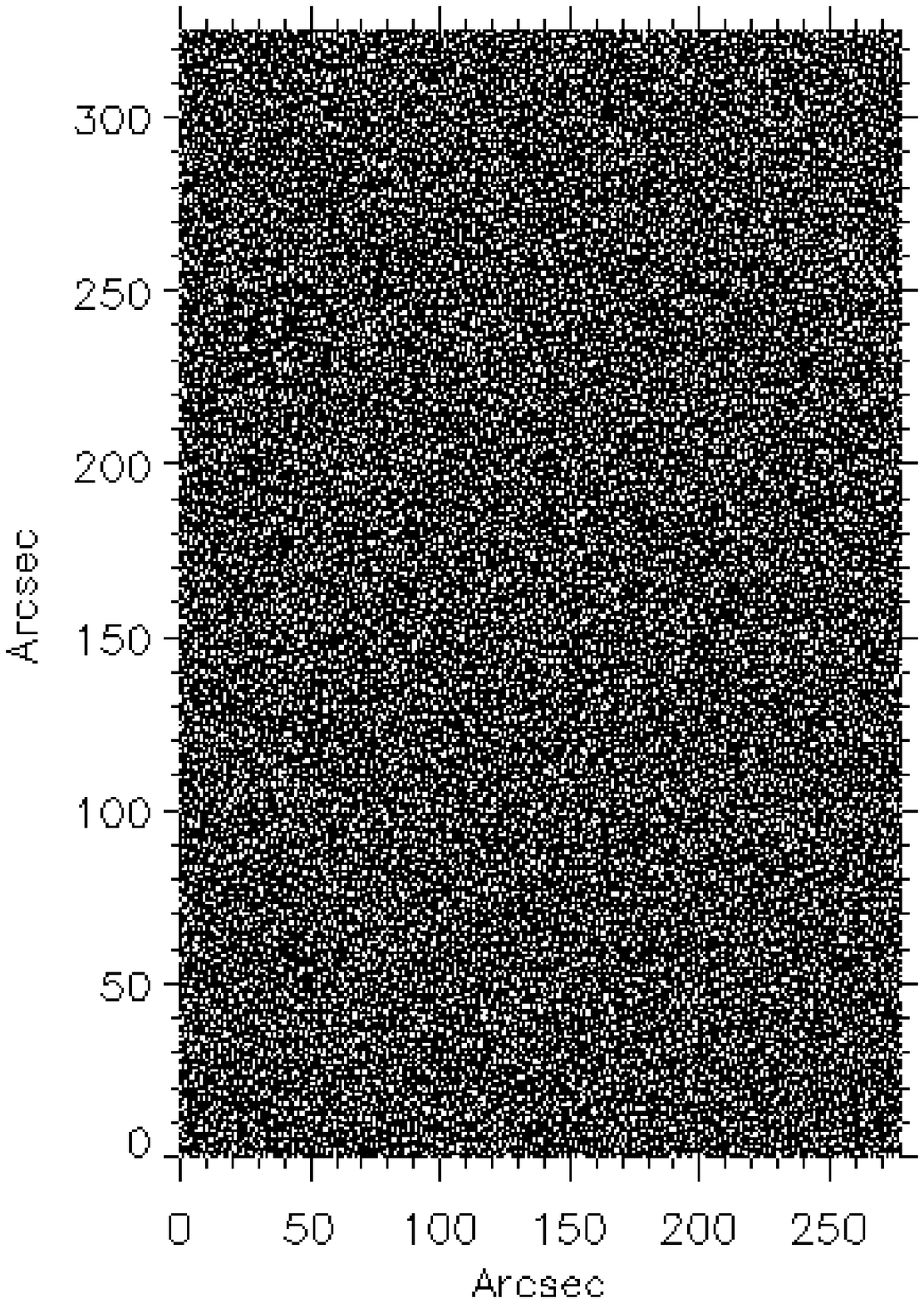}}\\
\centerline{\resizebox{6.2cm}{6.2cm}{\includegraphics{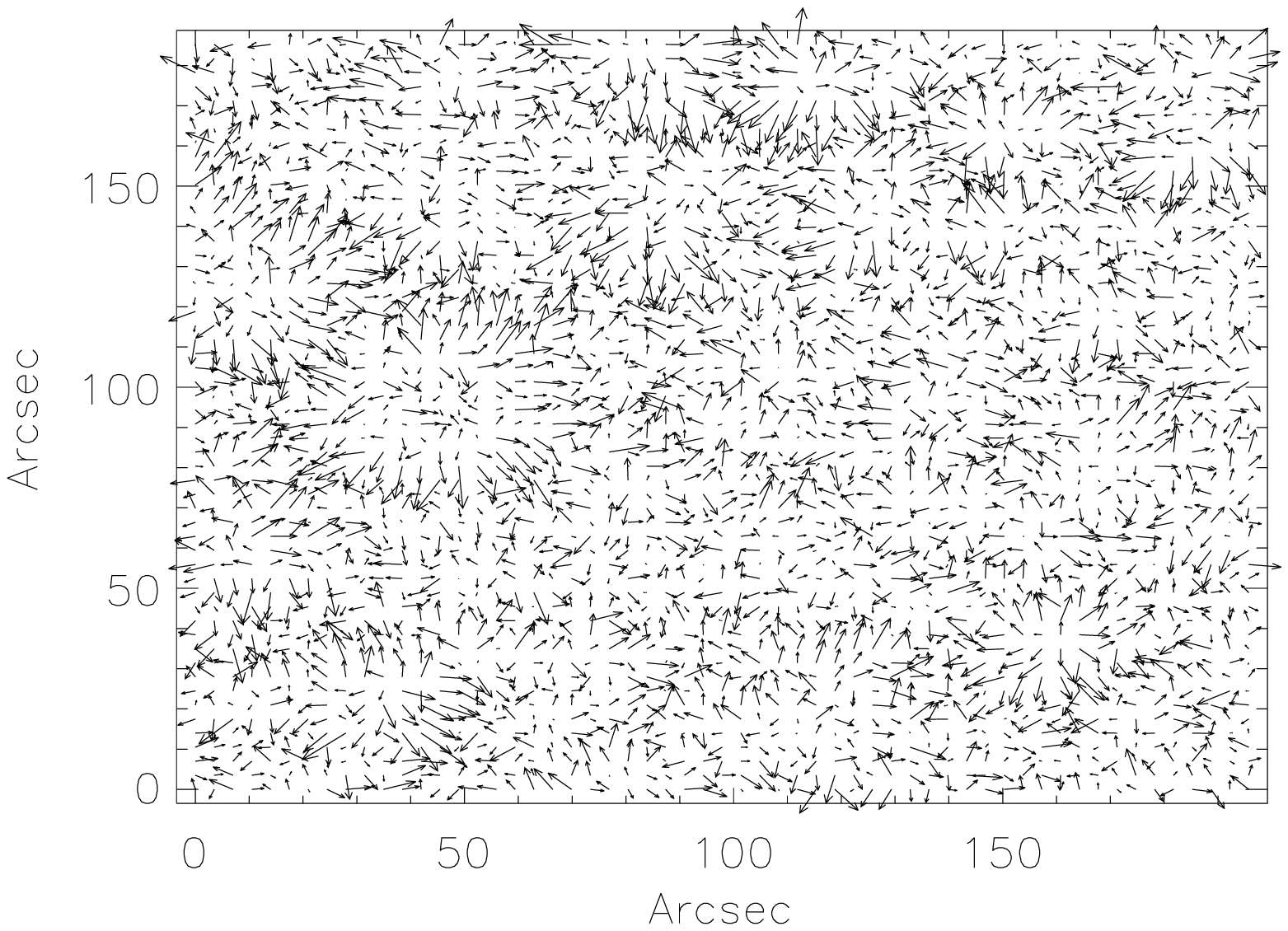}}}
\caption[]{Close-up view of the field around a spot as delineated in
Fig.~\ref{full_sun} (top left) with its segmentation (top right). Below,
resulting velocity field from granule tracking (note the clear signature
of the moat flow associated with the spot near x=50,y=100).}
\label{sunspot}
\end{figure*}

\subsection{Segmentation and granule identification}

To identify a granule one needs to establish a local criterion to decide
whether a given pixel belongs to a given structure.  This
criterion needs to be local to avoid threshold effects due to
large-scale variations of the intensity. The most efficient segmentation
algorithm for solar granulation is an adaptated version of  Strous'
segmentation (curvature-based criterion \cite{Strous94}) described in
\cite{RRRD07}. This image segmentation is very efficient at removing
large-scale intensity fluctuations. An example is shown in
Fig.~\ref{full_sun}. To summarize, it consists of the following steps:

\begin{itemize}
\item Calculation of the ``minimal curvature image'': for each pixel,
the minimal curvature among the four directions is computed.
\item Detection of the granules as non-negative curvature pixels in the
minimal curvature image.
\item Extension of the detected granules with points whose minimal
curvature value is higher than a given (negative) threshold,
while keeping a minimal distance of one pixel between each granule.
\end{itemize}

Once the image has been segmented, each granule needs to be identified.
This operation, called connected-component labeling, is an algorithmic
application of the graph theory, where subsets of connected components
are uniquely labeled based on a given heuristic. Connected-component
labeling is used to detect the connected regions in the binary digital
images produced by our segmentation algorithm.  In this labeling process,
a pixel belongs to a granule if it shares at least one side with another
pixel of the granule.

\begin{figure}
\centerline{\psfig{figure=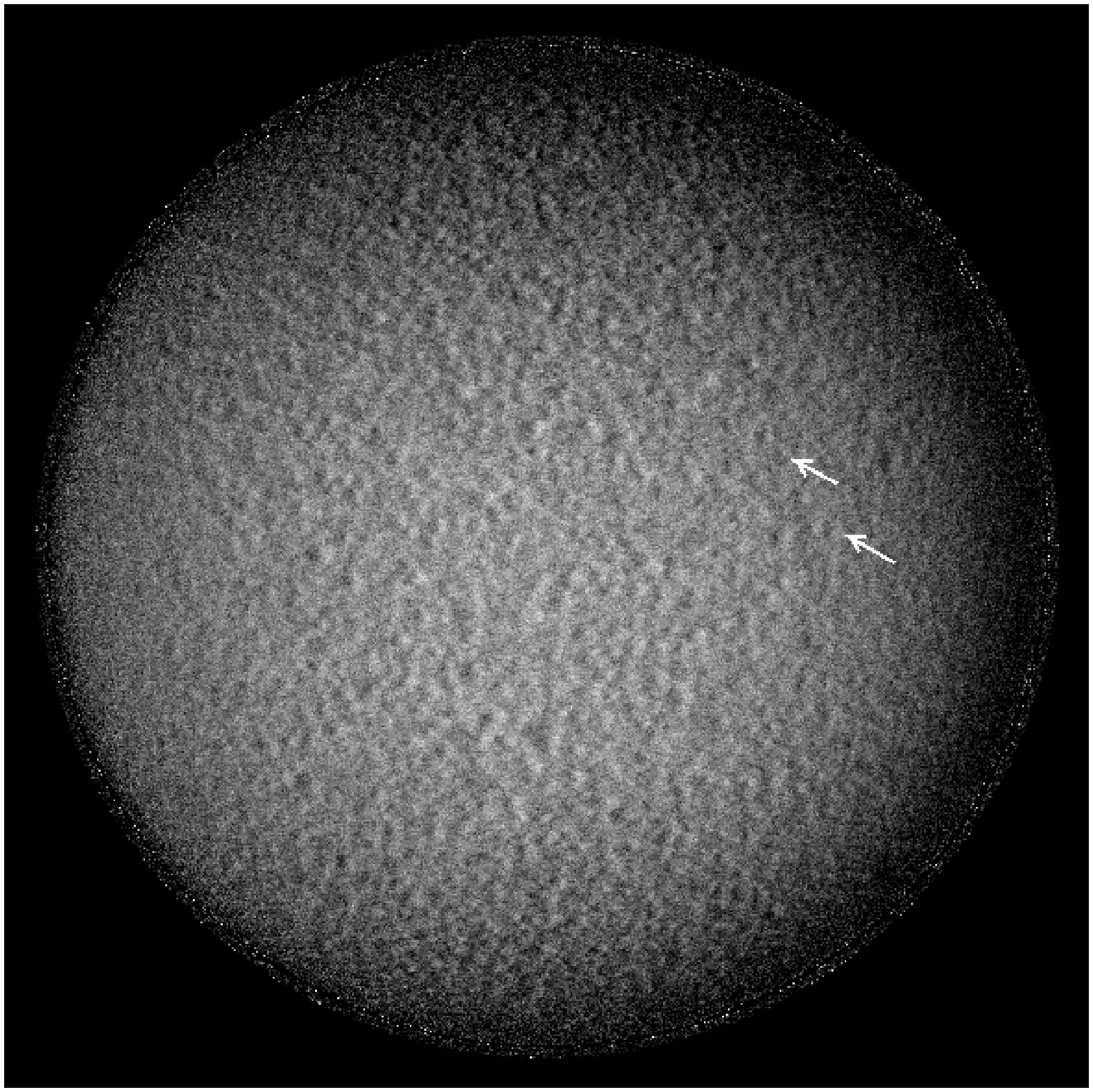,width=9 cm}}
\caption[]{Full Sun Vx component from a two-hour sequence, spatial resolution 2.5 Mm.
 Arrows indicate the location of supergranule visible also on the Doppler map. }
\label{Vx2h}
\end{figure}

\begin{figure}
\centerline{\psfig{figure=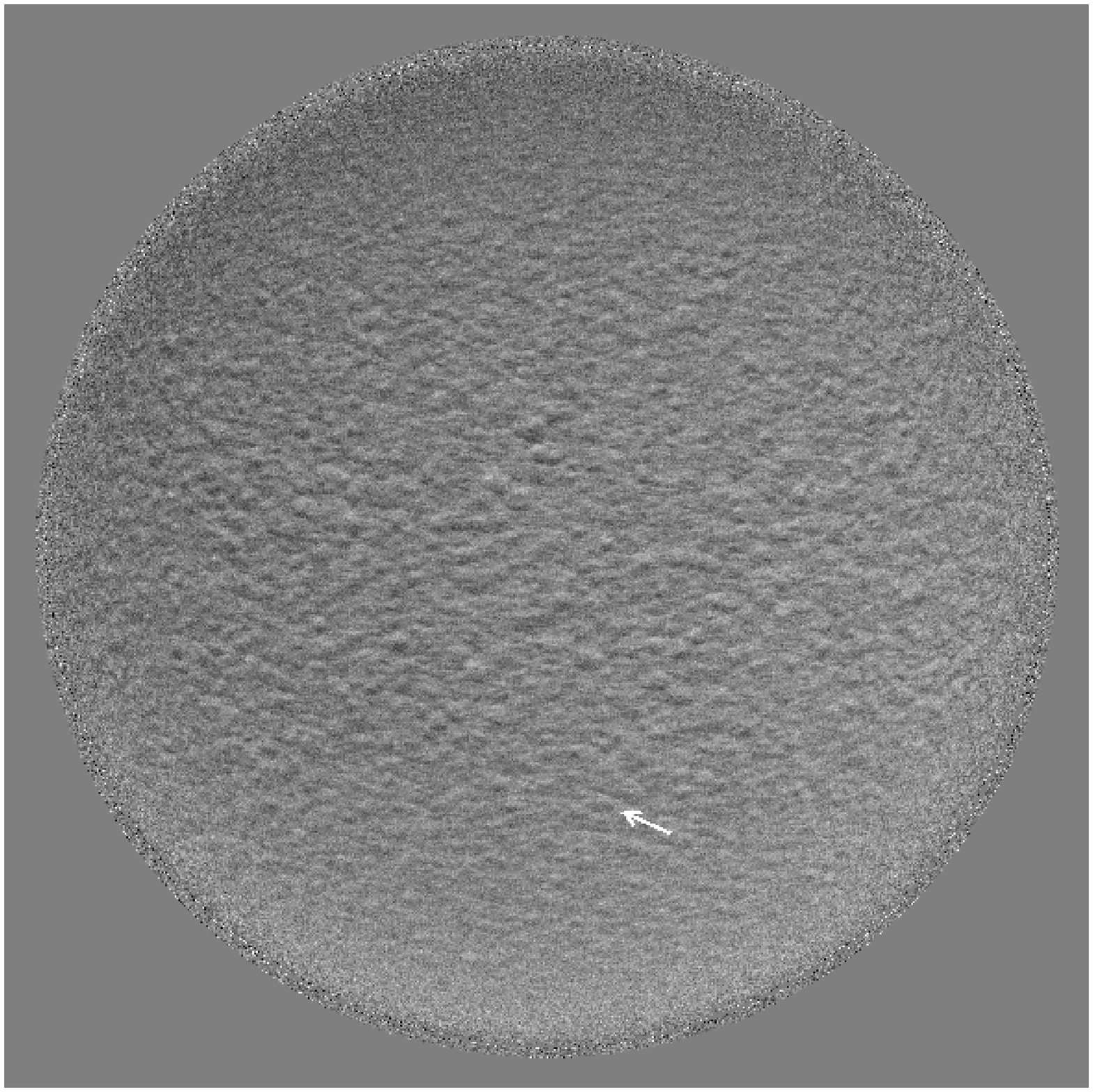,width=9 cm}}
\caption[]{Full Sun Vy component from two-hour sequence, spatial resolution 2.5 Mm.
 Arrow indicates the location a supergranule visible also on the Doppler map.}
\label{Vy2h}
\end{figure}
\begin{figure}
\centerline{\psfig{figure=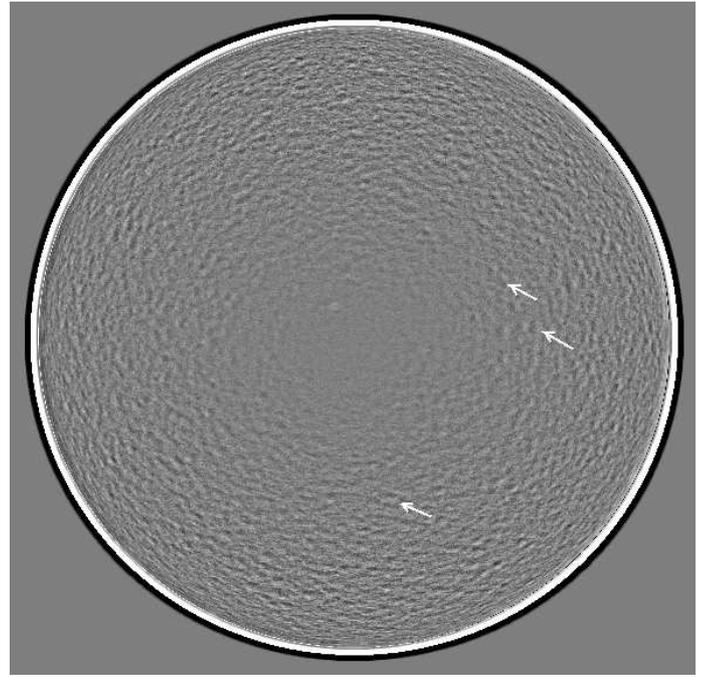,width=9 cm}}
\caption[]{Averaged Dopplergram from two-hour sequence where the solar rotation has been removed.
  Arrows indicate the location of supergranule.}
\label{Doppler}
\end{figure}

\begin{figure}
\centerline{\psfig{figure=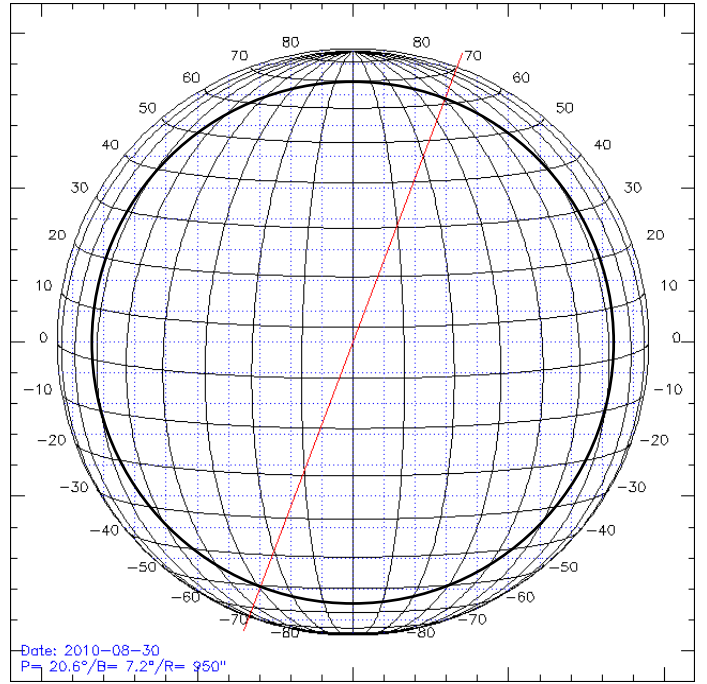,width=9 cm}}
\caption[]{ Solar grid where the dark circle indicates the limit of the validity 
 of the horizontal velocity measurement.  }
\label{Grid}
\end{figure}

\subsection{Measuring the velocities}

Once the granules have been labeled in each frame, their trajectories
are identified by comparing two consecutive images.  During their
lifetime, granules can split or merge into multiple objects. Disappearance
or appearance between two successive frames must be taken into
account. Then, the life of coherent objects (\textit{i.e}. granules)
is defined between its appearance and disappearance if the granule does
not split or merge.  When the granule splits, the life of granule is
stopped and its ``children" are considered as new granules. In the same
way, when granules merge, the lives of the granules that merge are
stopped and the new granule issued from the merging is considered as
a new granule. Thus, we can follow a coherent structure between their
birth and death. In the previous CST version, the temporal labeling of the 
coherent structure was performed from one frame to the next by following
the barycenter trajectory \citep{RRMV99,RRRD07}  and was essentially
applied to the aligned data where large-scale motions were removed (solar
rotation, drifts, etc.). However, when this method is applied to the
HMI/SDO data, it  generates some misidentifications of the barycenters
between close granules because of the high value of the rotation
velocity (around 2 km~s$^{-1}$). To avoid this extra noise, we now performed
a time labeling of granules, that takes into account a common granule
area between image $t$ and $t+1$. This method ensures the
temporal link between two snapshots of an evolving granule and improves
their temporal labeling.

Hence, each coherent structure (i) is defined by six values: 

\begin{enumerate}
\item birth time $Tb_i$  and death time $Td_i$ 
 
\item $(X_i,Y_i)_b$ and $(X_i,Y_i)_d$ the positions at time $Tb_i$ and $Td_i$

\item $V_x$ and $V_y$  are in the heliocentric-cartesian coordinates
\citep{Thompson2006}, respectively the horizontal component of velocity
in $x$  ($x$ increasing toward solar west) and in y (increasing toward
solar north).

  \end{enumerate}

From the birth and death locations and lifetimes of each coherent
structure, we derived a trajectory and mean velocity.
To reduce the noise, coherent structures with a lifetime shorter than 180 seconds 
were ignored.

If one analyzes a long times series of images, it is useful to
determine the time evolution of the velocity field; for this purpose a
time window lasting $\delta_t$ was used and trajectories were
restricted to the time window.  Hence, for a given time window, we derived
a set of trajectories and velocities. The values of the velocities
are of course not uniformly distributed in the field of view, and we
need to know how they constrain the velocity field at a given scale:
small-scale components are weakly constrained while large-scale ones
are highly constrained. The maximum resolution for the velocity field
is given by the density of trajectories. As pointed out in \cite{RRLNS01},
granules cannot be used to trace plasma flow below a scale of 2.5~Mm
(except for very rapid flows like in an ``explosion'' of granules); thus
the determination of a large-scale flow needs a mesh size not smaller
than 1.25~Mm.

\cite{RRRD07} indicated that the minimum size of the velocity mesh grid is
related to the temporal resolution. They showed that  one typical granule
trajectory occupies a ``volume" of 1200Mm$^2$s. When the ``space-time"
resolution does not reach this limit, many granules contribute to the
velocity in one mesh point. Their mean velocity is considered as the
true local velocity but local fluctuations around this may also give some
information on the local strength of convection. Because granules do not 
sample the field of view uniformly, a reconstruction
of the velocity field along with its derivatives such as the divergence
\mbox{$D=\partial_x V_x+\partial_y V_y$} or the $z$-component of  vorticity
\mbox{$\zeta=\partial_x V_y-\partial_y V_x$} requires some interpolation.
Like \cite{RRRD07} we used the multi-resolution
analysis for the interpolation, which limits the effects of noise and error propagation.

The smallest  spatial and temporal resolutions achieved in the
horizontal velocity map produced using the new CST algorithm are 2.5 Mm and 30
min, respectively.

Fig.~\ref{sunspot} shows some details around a sunspot in the
intensity field and its segmentation where granules are visible;
we plot an enlargement of the flow fields around the sunspot where divergent 
structures and the sunspot moat are clearly visible.

On the full Sun white-light observation around $5\times 10^5$ granules are
detected. During a time tracking of 30 minutes around $2.2 \times 10^6$
coherent structures are identified. This number increases up to $7.5
\times 10^6$ for a time sequence of two-hours.
We show an example of  the full Sun $V_x$ and $V_y$  in Fig.~\ref{Vx2h} and ~\ref{Vy2h}  
components from a two-hour sequence with a spatial resolution of 2.5 Mm. Arrows indicate
the location of supergranules that are visible also on the Doppler map 
(Fig.~\ref{Doppler}). Owing to the radial flows within supergranules,  two
supergranules are marked in the Vx map close to the western limb where they are
easier to identify than in the east-west component of the flow (and vice versa for the Vy map).
 The circle inside the solar grid in Fig.~\ref{Grid} represents the limit 
up to which the projected Vx and Vy are correctly measured. Beyond this circle, granule 
tracking is difficult and errors increase rapidly as we near the limb. The useful 
field lies in the latitude and longitude range between $\pm60\degr$ in 
latitude and longitude for a two-hour sequence. Because this field covers a large part of the 
visible Sun area (80\%), we define it as a quasi full-disk map of photospheric horizontal velocities.

\section{Precision }

\begin{figure}
\resizebox{4cm}{4cm}{\includegraphics{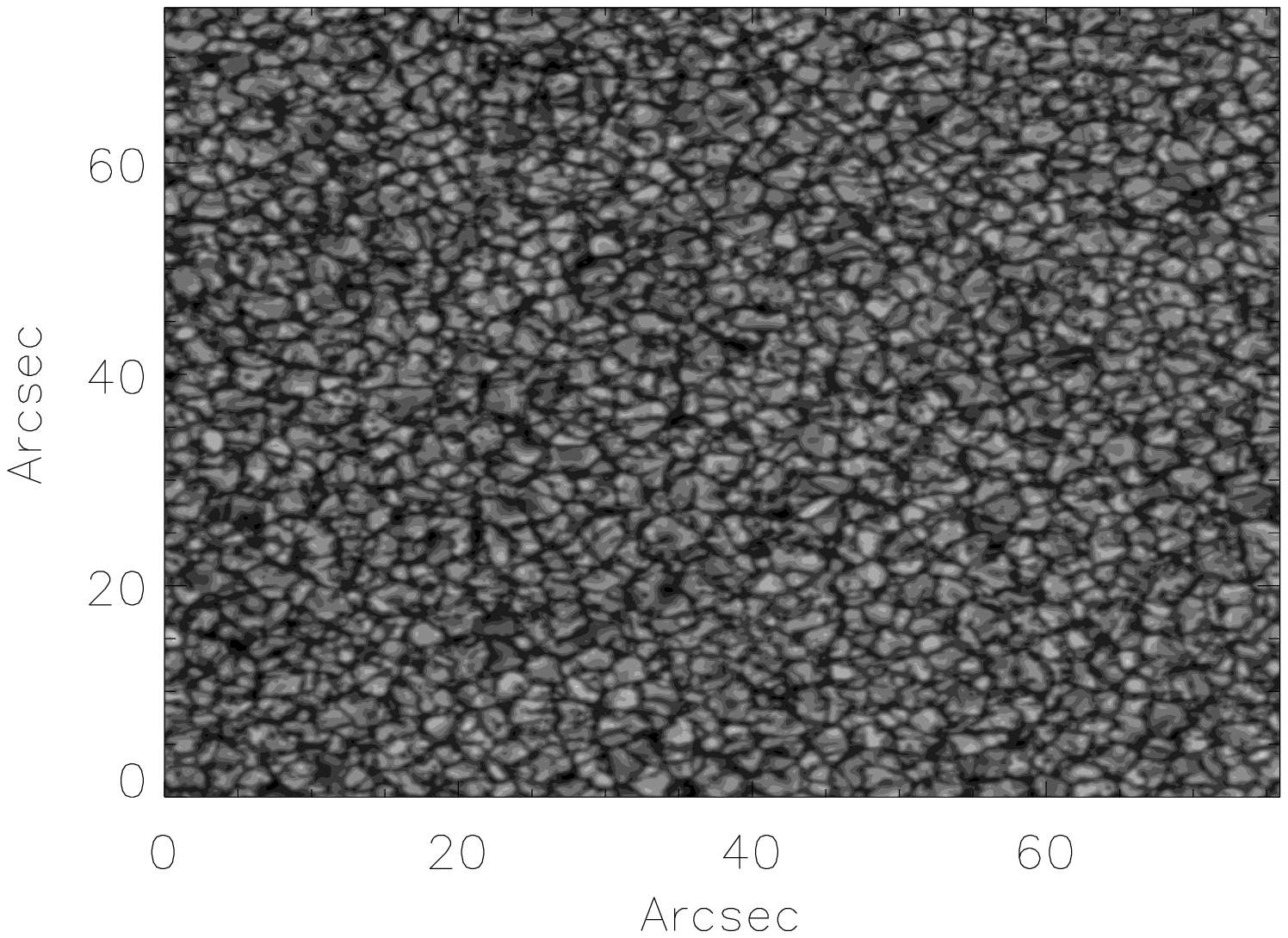}}
\resizebox{4cm}{4cm}{\includegraphics{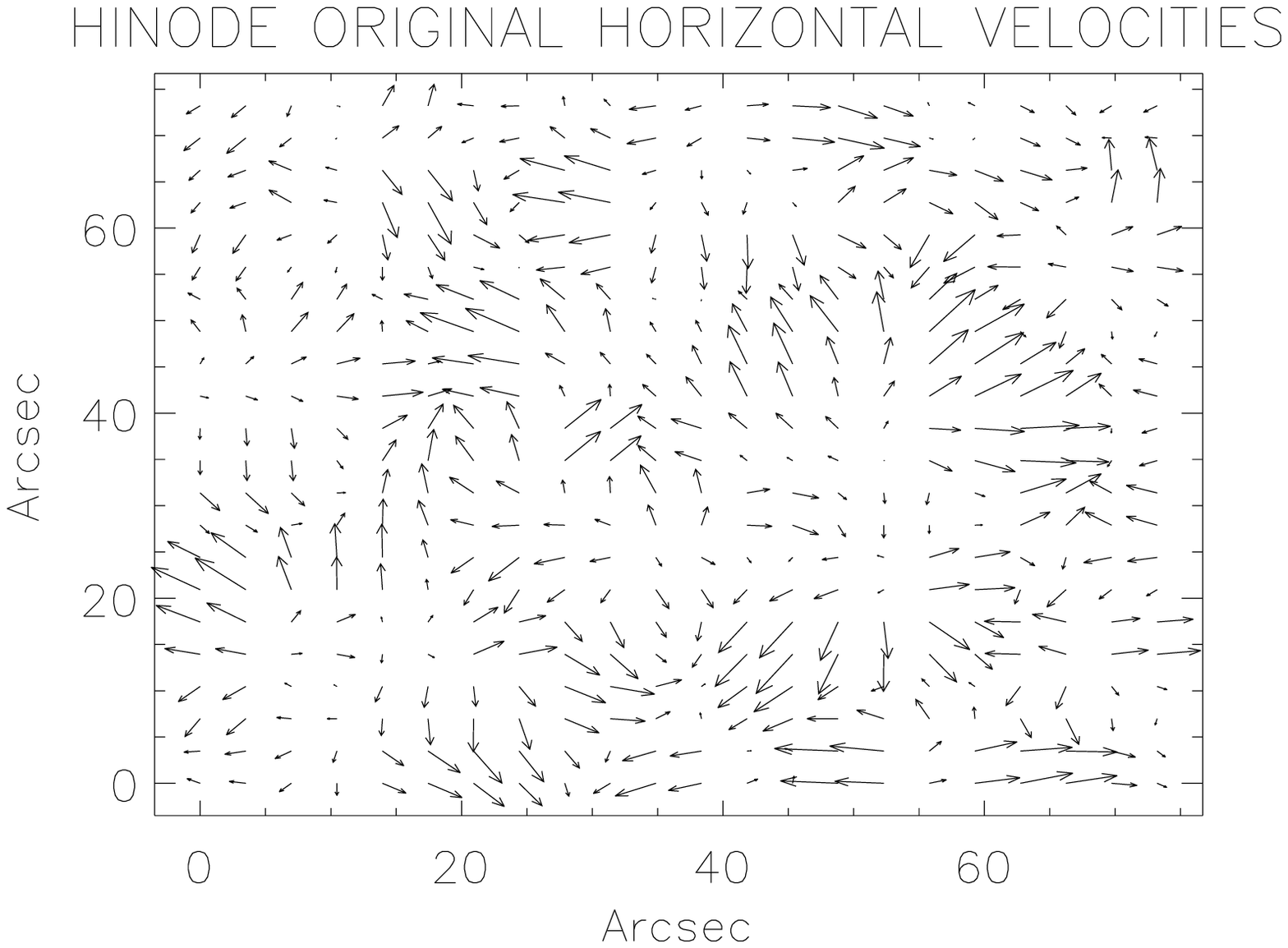}}\\
\resizebox{4cm}{4cm}{\includegraphics{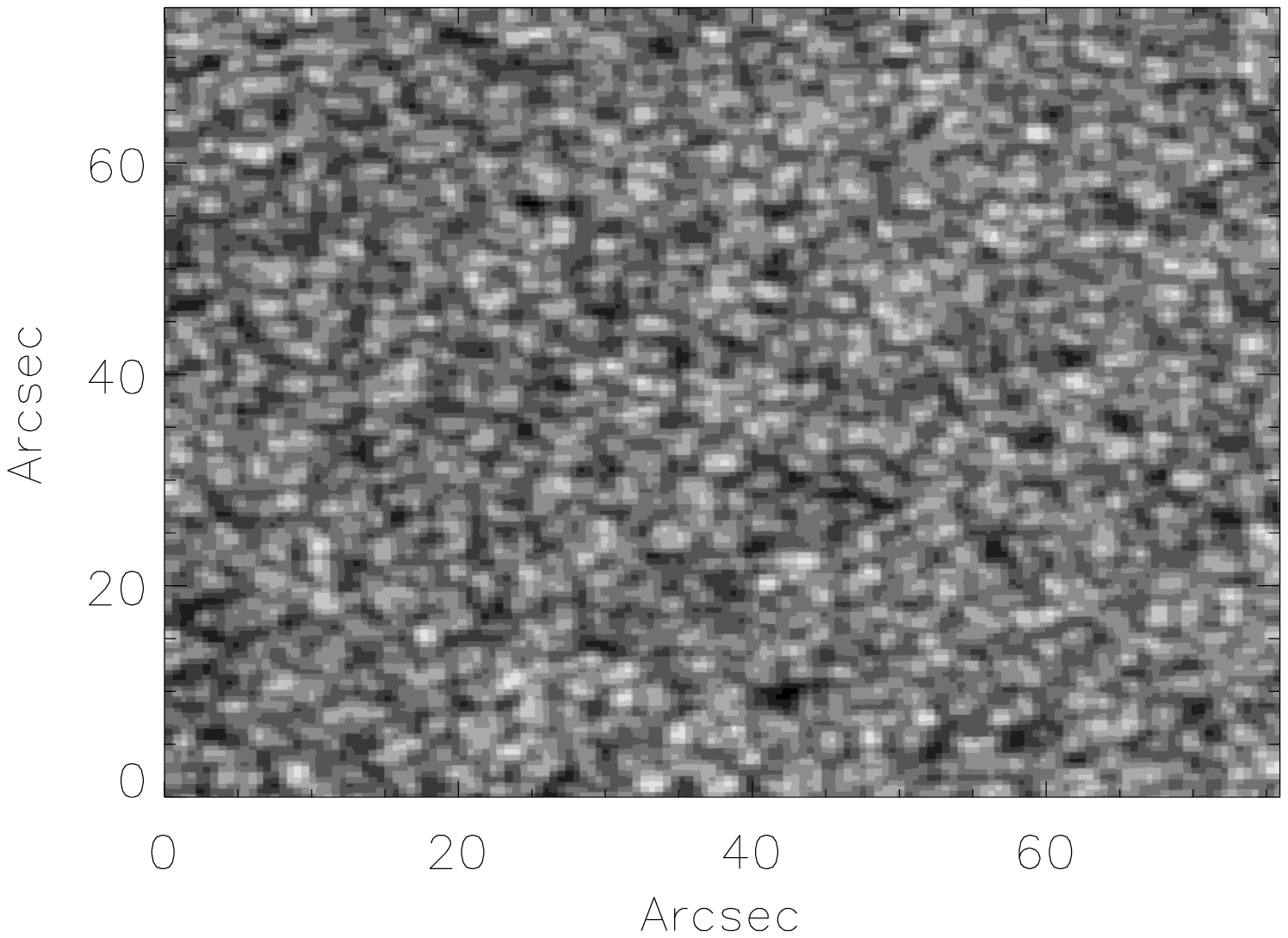}}
\resizebox{4cm}{4cm}{\includegraphics{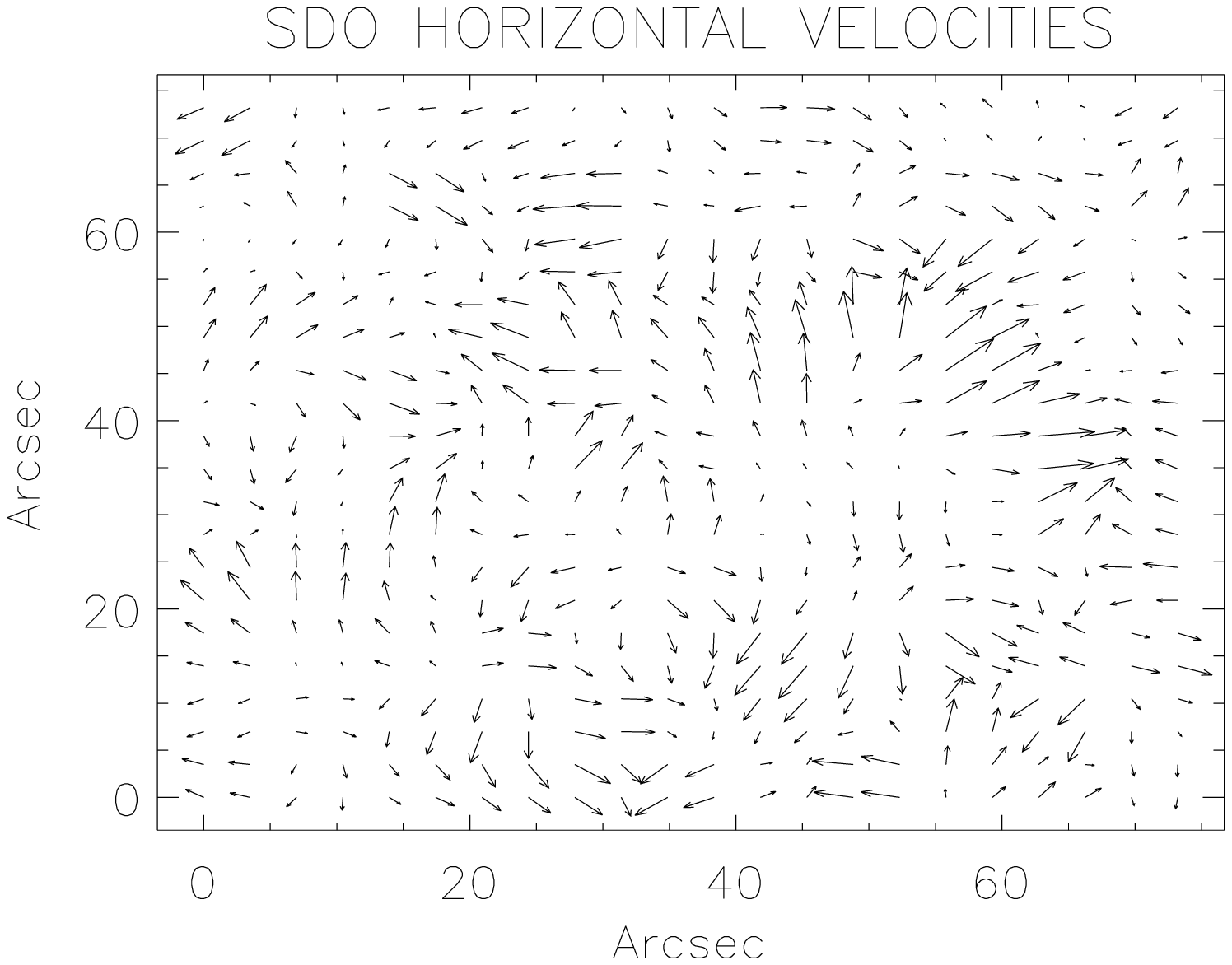}}\\
\resizebox{4cm}{4cm}{\includegraphics{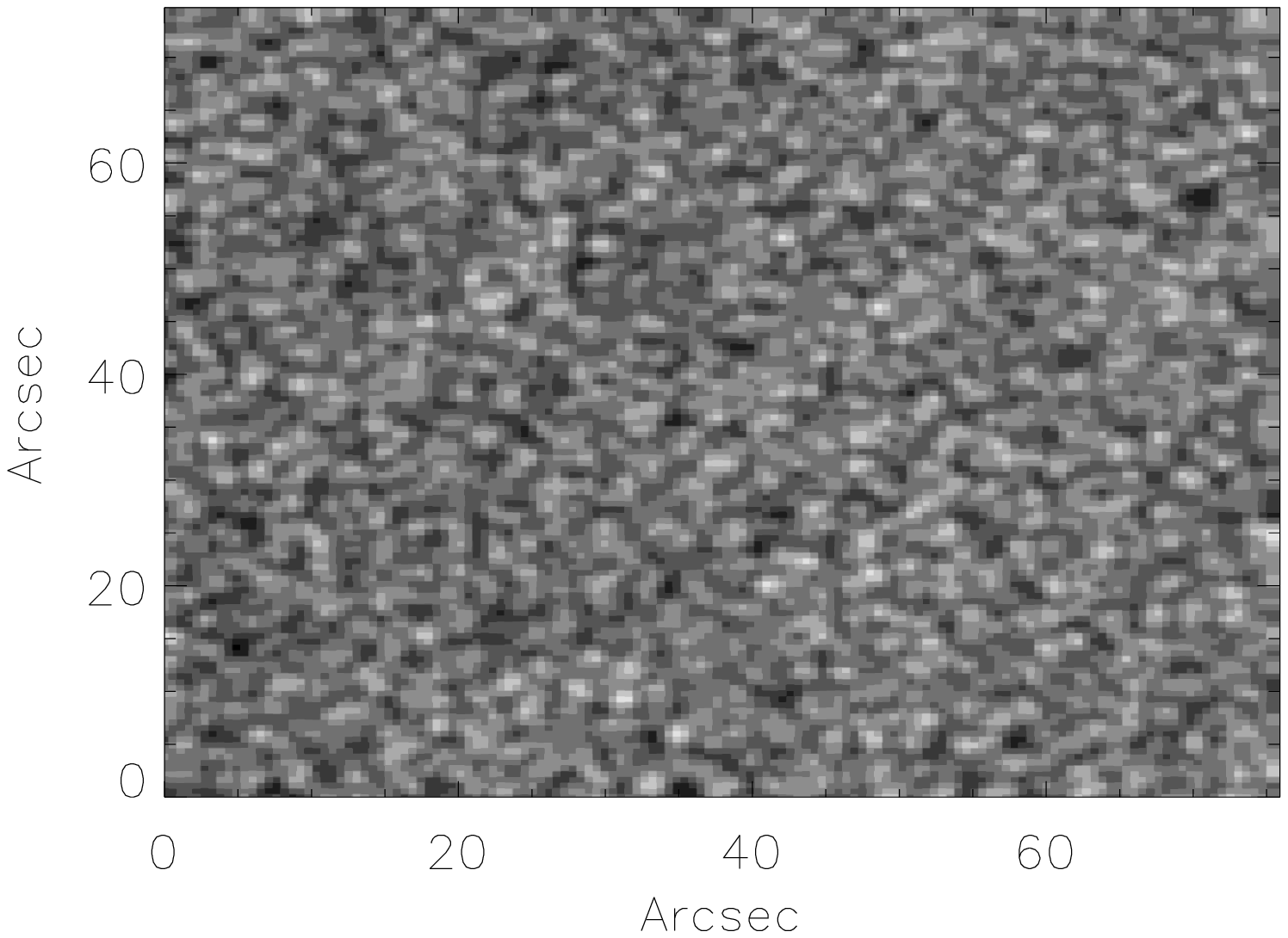}}
\resizebox{4cm}{4cm}{\includegraphics{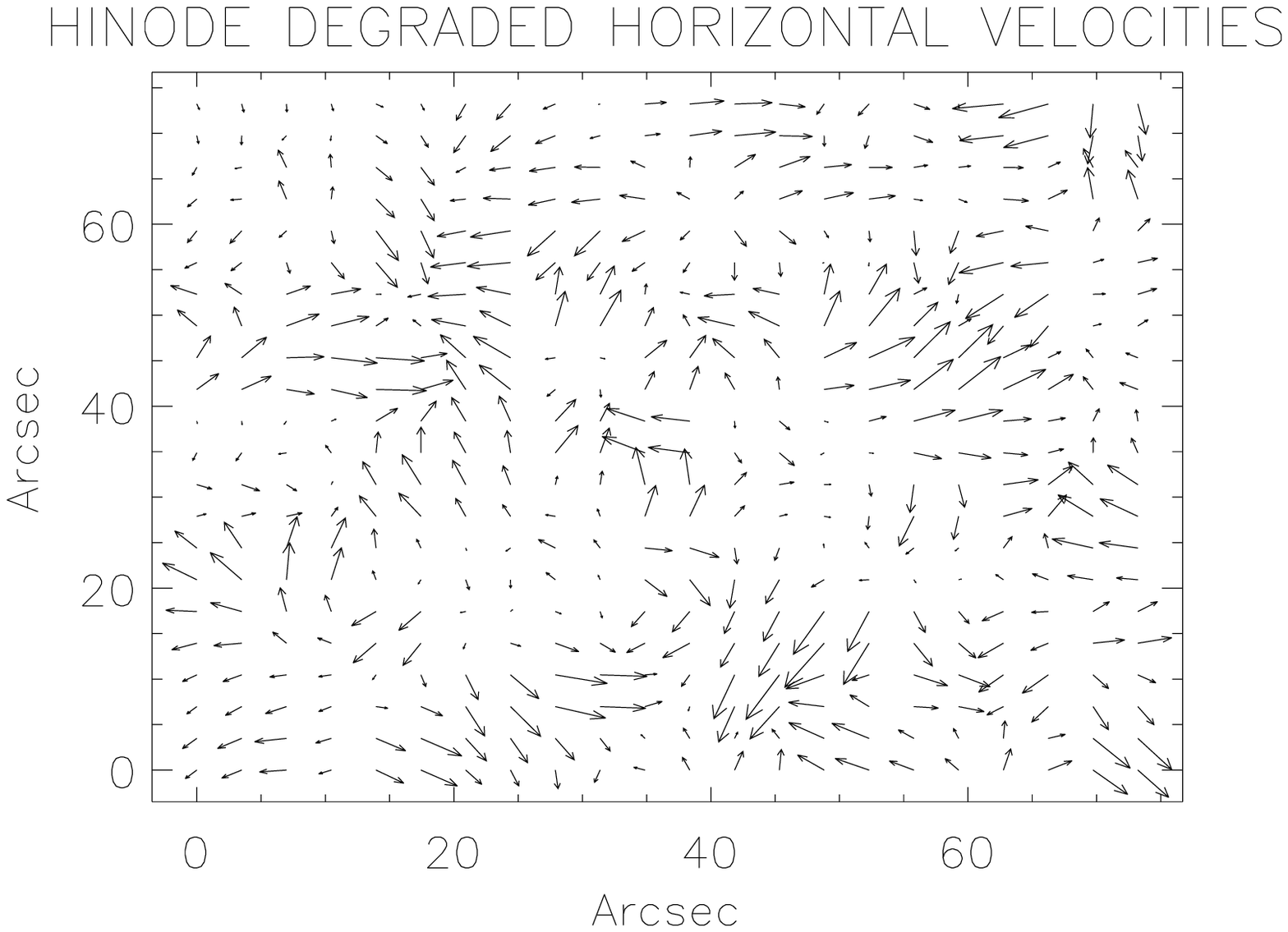}}\\
\resizebox{4cm}{4cm}{\includegraphics{fig/18678f7a.ps}}
\resizebox{4cm}{4cm}{\includegraphics{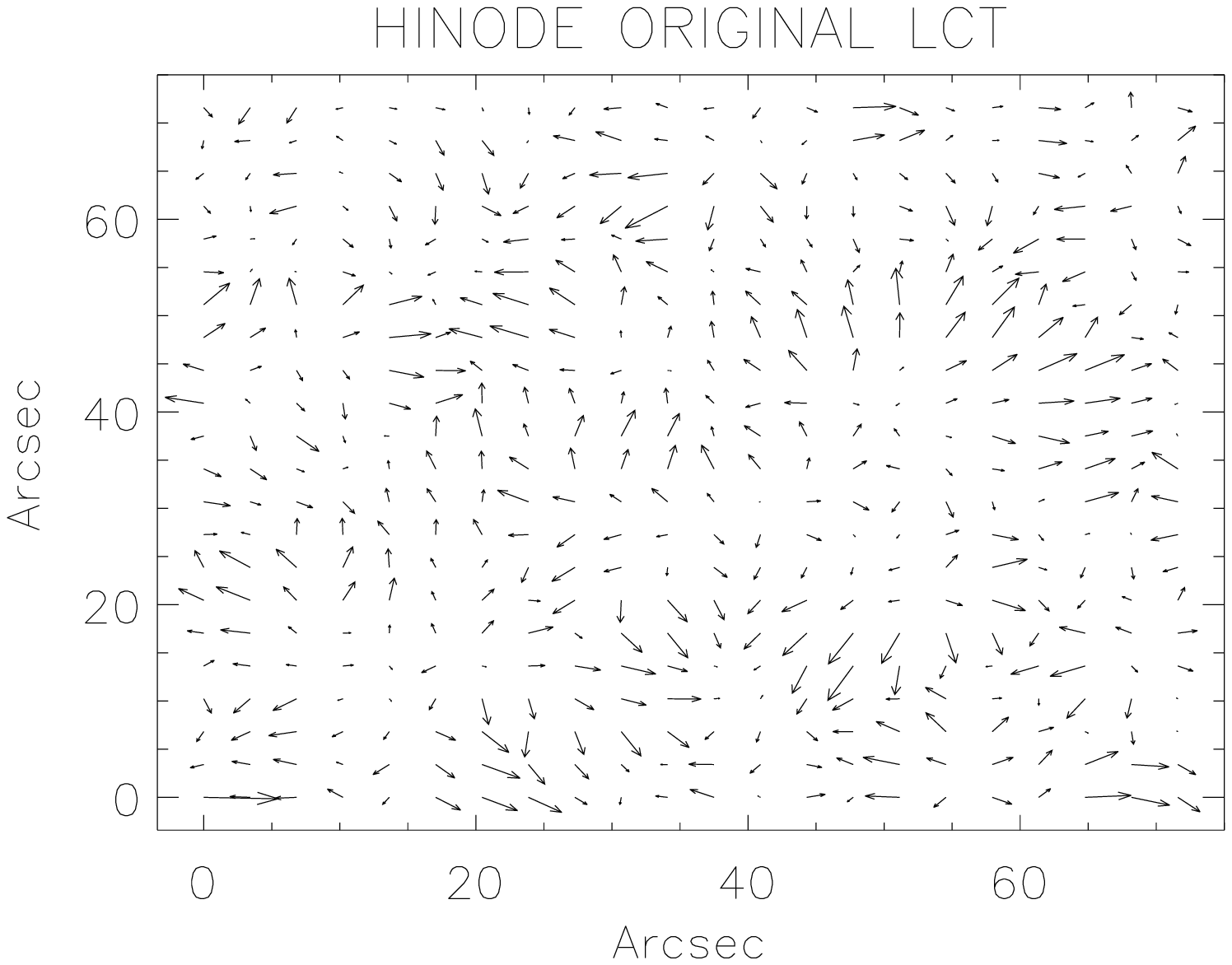}}

\caption[]{ Solar granulation from \textit{Hinode} (1 pixel =
$0\farcs1089$) on  August, 30 2010 8h04mn33s (top left) and   horizontal
velocities from a one-hour sequence (top right).  Solar granulation from
SDO (1 pixel = $0\farcs5042$ ), on  August, 30 2010 8h04mn30s (middle
left) and horizontal velocities from a one-hour sequence (middle right).
Solar granulation from \textit{Hinode} degraded to a pixel of $0\farcs5$
on  August, 30 2010  8h04mn30s (bottom left) and horizontal velocities
from a one-hour sequence (bottom right).}

\label{compar}
\end{figure}

\begin{figure}
\centerline{\psfig{figure=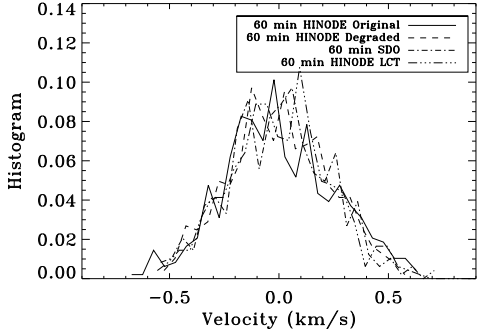,width=7 cm}}
\caption[]{Velocity  histograms from CST and LCT are  plotted in Fig.~\ref{compar} }
\label{histo_velo}
\end{figure}

Following \cite{TRMR07}, we can estimate the error propagation on the velocity
of the CST algorithm. Over an image, granules are much less dense
than pixels, therefore the velocity field is sampled on a much coarser grid
whose elements are of size $\delta$ here 2.5 Mm. The velocity at a grid
coordinate $(x,y)$ is assumed to be the average of the velocity of
granules whose average coordinates  belong to the domain $D$  around
$(x,y)$ (i.e. in $[x-\delta/2,x+\delta/2],[y-\delta/2,y+\delta/2]$
). The spatial resolution of the velocity field is given by the mesh
size $\delta$.  If we assume that the dispersion on the barycenter is
the same for the whole time series ($\sigma$), and that the time interval
$\Delta t_k$ (of the granule $k$) is the same for all granules (i.e. all
granules have the same lifetime $\Delta t$), then the expression of the
dispersion of the average velocity  \citep{TRMR07} is

\begin{equation}
\delta V=\frac{\sqrt{2}\sigma}{\sqrt{N}\Delta t}.\label{preci}
\end{equation}
where $N$ is the number of trajectories falling in $D$.
We found for the full Sun white-light SDO data for a time
sequence of 30 minutes and a mesh size of 2.5 Mm, that N=8. The precision
of the location of the barycenter of granule is estimated to half a
pixel, which represents $0\farcs25$ (183 km). With a median lifetime of
405 seconds $\delta V=0.22$ km~s$^{-1}$,  which is good enough at a scale of 2500~km,
\textit{i.e}. at the scale where granule motion  traces the large-scale
plasma flows. \cite{TRMR07} showed  that precise
velocity values need many granules in a grid element and a long-time
interval. In other words, errors are less if a coarse resolution in
space and time is used.

When moving to the limb, the detection of granules becomes more difficult
but the mesh size $\delta$ covers a larger area on the Sun's surface. Both phenomena
more or less cancel each other out and we find that close to the limb N=7 and
$\delta V=0.24$ km~s$^{-1}$.

Accordingly, we estimate that  an error of  $\delta V=0.25$ km~s$^{-1}$  for the
HMI/SDO data for a time sequence of 30 min is acceptable since most of
the granules are showing velocities in the range 0.3--1.8km~s$^{-1}$ in
our analysis. This error seems to be large, but this is due in great 
part to the pixel size $0\farcs5$, which allows us to follow only the large granules
on the Sun surface. However, when the velocities are averaged in space and time 
this error decreases for example down to 0.06~km~s$^{-1}$ at the disk center with a 
window of  $20~\arcsec\times20~\arcsec$ for a 24-h sequence (see below).
  This estimate error of the smoothed data is very similar to the one deduced 
from  helioseismology analysis, which is found to be 30 m/s for one day data (Table 2 
of \cite{svanda11} ).

\section{Comparison of HINODE and SDO velocity fields}

To determine the quality of the velocities measured with white-light 
HMI/SDO data, quasi simultaneous observations of the Sun's surface
with very high spatial resolution were obtained with the \textit{Hinode}
satellite during three-hours. The data were carefully aligned, k-$\omega$
filtered and resized at the same scale (1 pixel =$0\farcs1089$). The 
\textit{Hinode} data were also degraded to the spatial resolution of
HMI/SDO for a detailed comparison.  Three temporal sequences were treated
exactly in the same way.  We applied the CST to these sequences and show the
results in Fig.~\ref{compar}. The correlation between velocities
of the different sequences is about 85\% and one can easily recognize common 
patterns between the time series. The histograms of the velocity in Fig.~\ref{histo_velo}
are very similar for all sequence, indicating that the amplitudes are
also correctly measured.
 For comparison, local correlation tracking (LCT) was applied to the same 
one-hour sequence. This flow field is shown in the bottom of Fig.~\ref{compar}.
Here we also find a good  correlation (about 75\%) between velocities computed 
by CST and LCT. The LCT velocities are lower, however than the CST velocities by 15\%, 
likely because of the spatial window size used in the LCT to convolve the data \citep{RRMV99}.

\begin{figure*}
\resizebox{8cm}{!}{\includegraphics{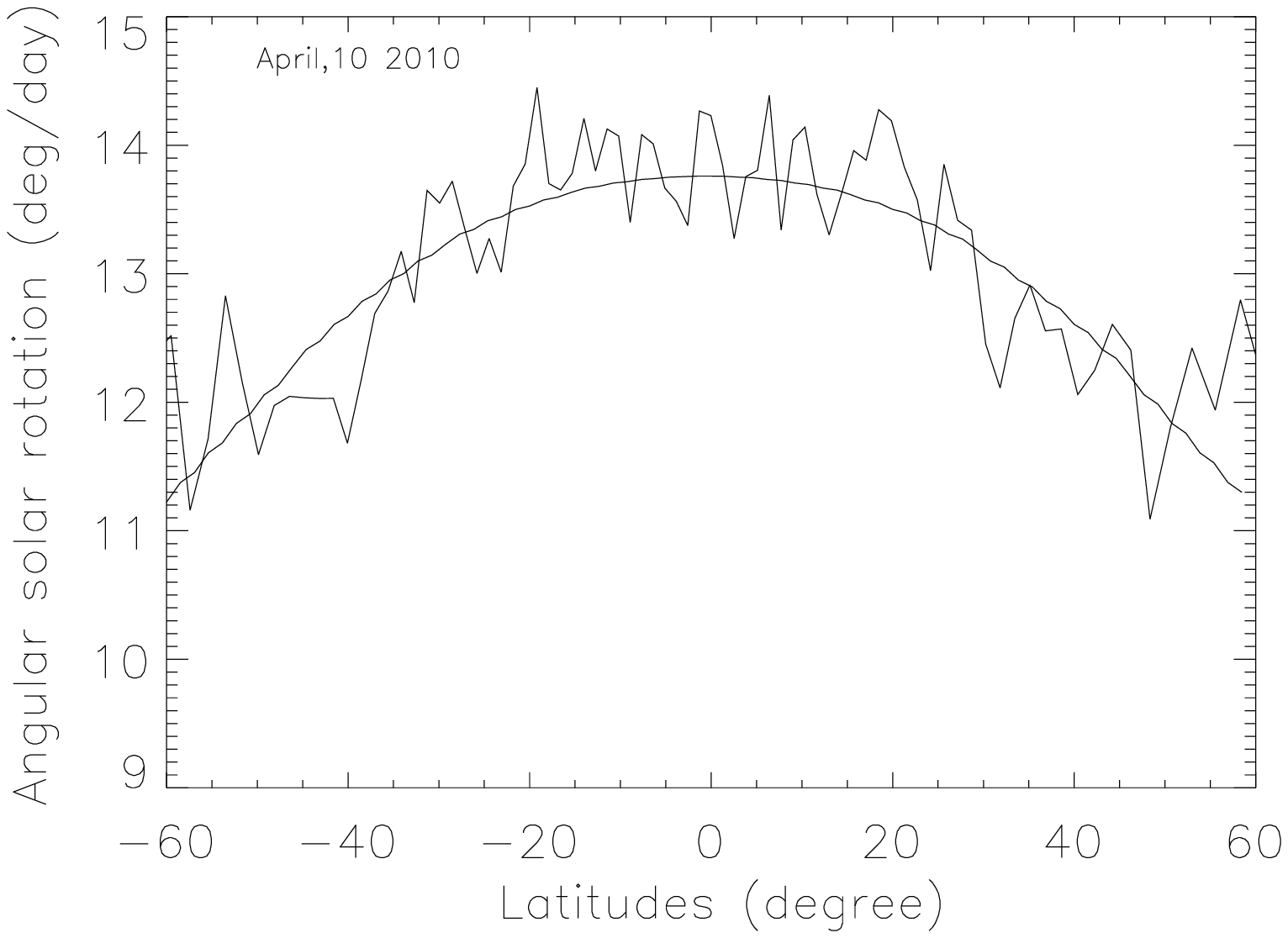}}
\resizebox{8cm}{!}{\includegraphics{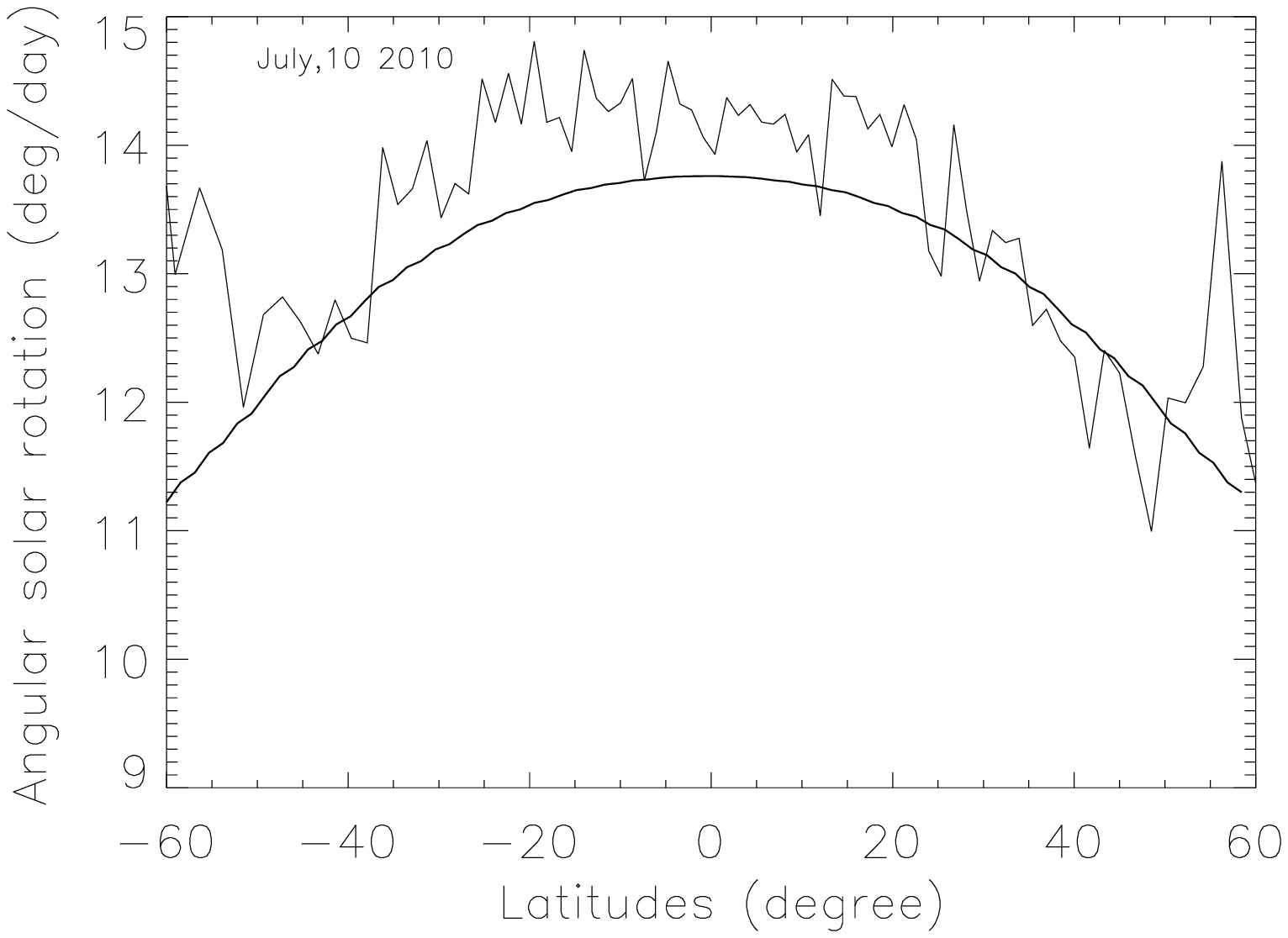}}\\
\resizebox{8cm}{!}{\includegraphics{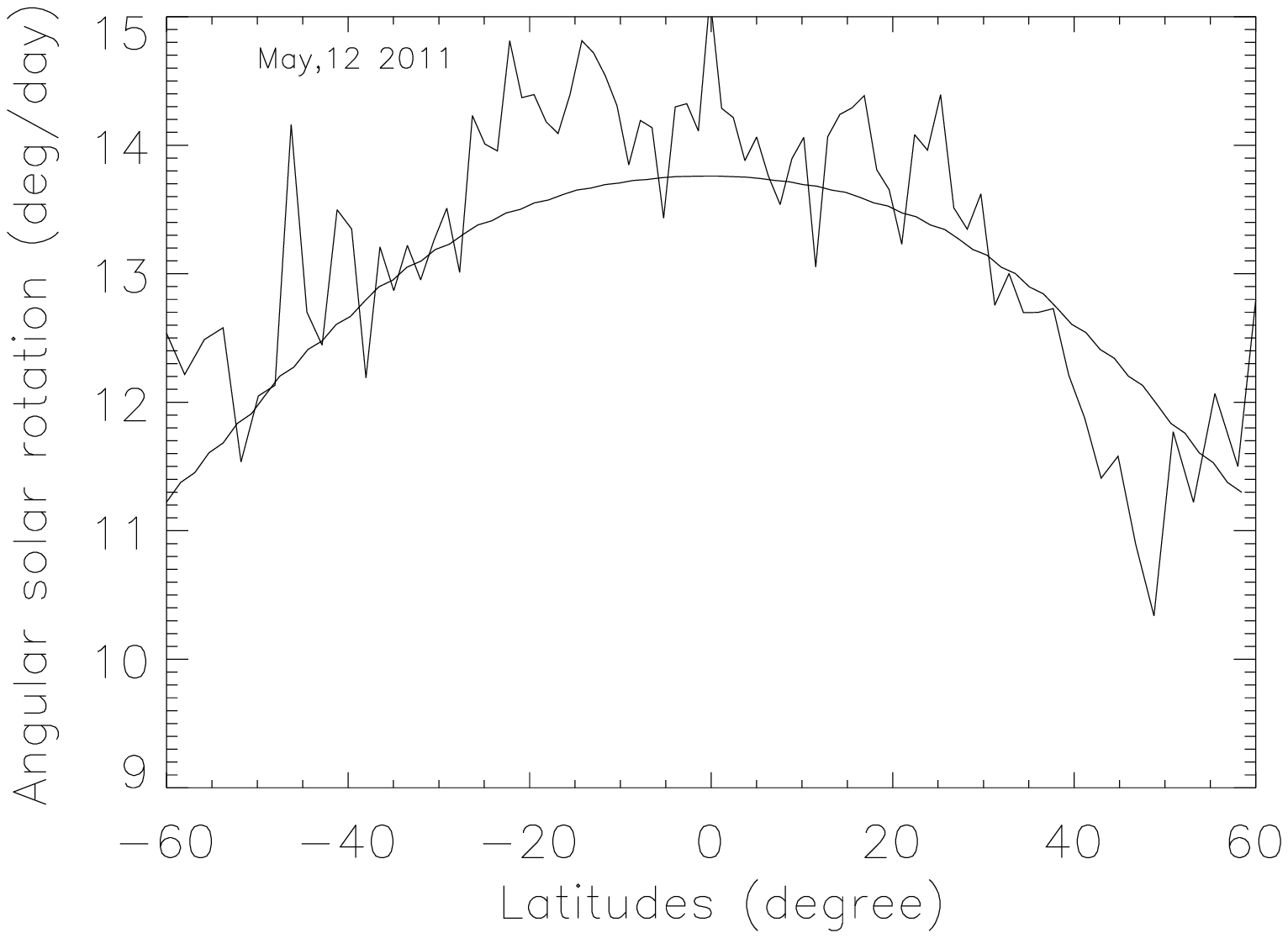}}
\resizebox{8cm}{!}{\includegraphics{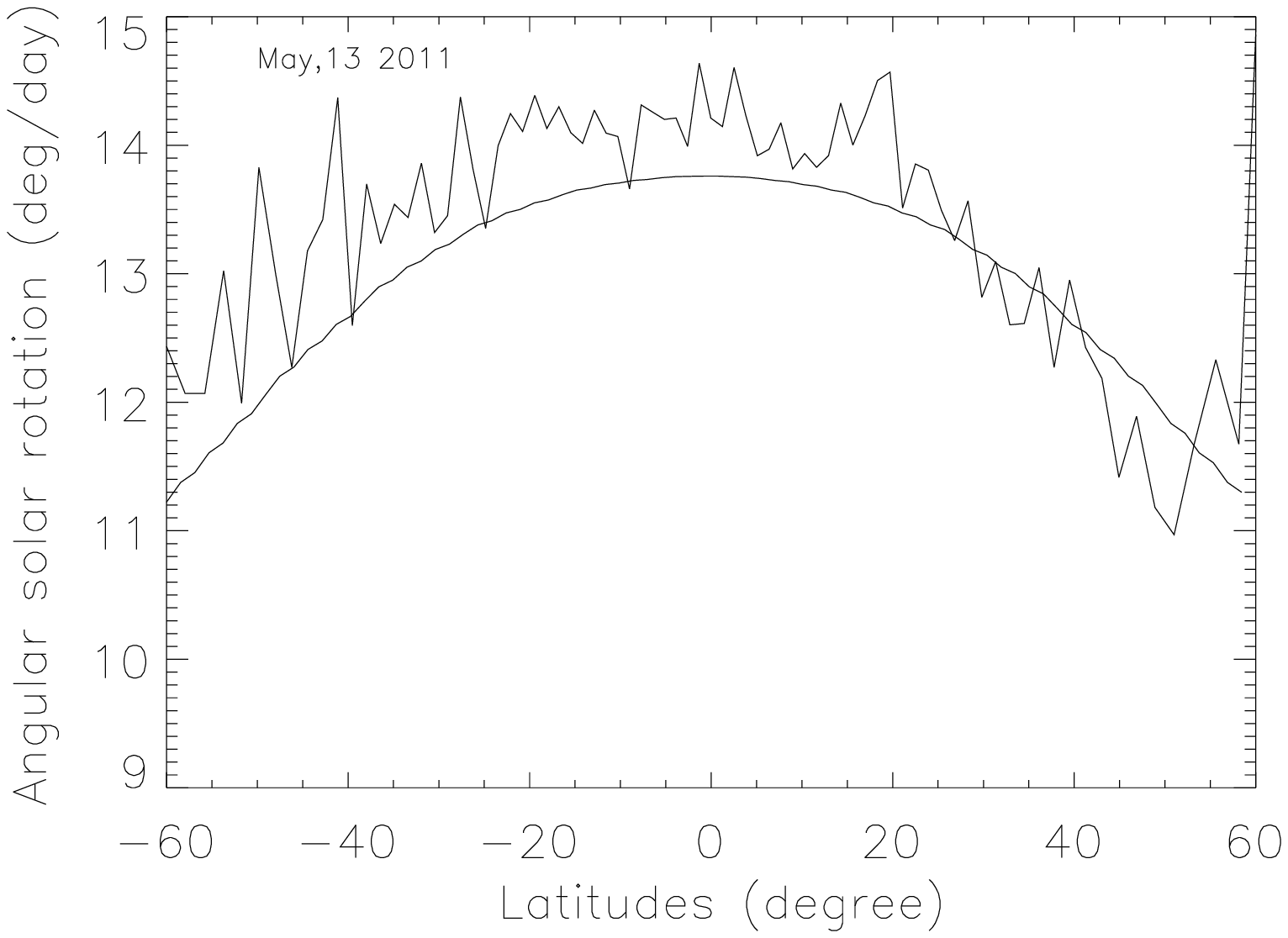}}\\
\caption[]{Solar differential rotation on April 10, 2010, July 10, 2010,
May 12, 2011, and May 13, 2011. We plot the rotation law determined by spectroscopic 
method \citep{HH70} in the four panels $\omega = 13.76 - 1.74 \times \sin^2\lambda - 2.19 \times \sin^4\lambda$
($\lambda$= latitude, $\omega$ = angular velocity in deg/day), giving an
equatorial velocity of 1.93 km~s$^{-1}$}

\label{rotdif}
\end{figure*}

The high correlation  between velocities and the quasi-identical amplitudes
of the velocities indicate a good agreement between velocities measured
with \textit{Hinode} data (high spatial resolution) and HMI/SDO data
(low spatial resolution). This allows us to use HMI/SDO data with some
confidence in determining the horizontal flows across the Sun.

\section{Power spectrum of the solar supergranulation}

 The kinetic energy distribution among the scales is represented
by the power spectrum of the velocity field. Like \cite{RMRRBP2008}, we computed  
the kinetic energy spectral density $E(k)$ for a time window of 120~min 
and spatial field of $650~\arcsec\times650~\arcsec$  centered on the disk 
center on August 30, 2010.

\begin{figure}
\centerline{\psfig{figure=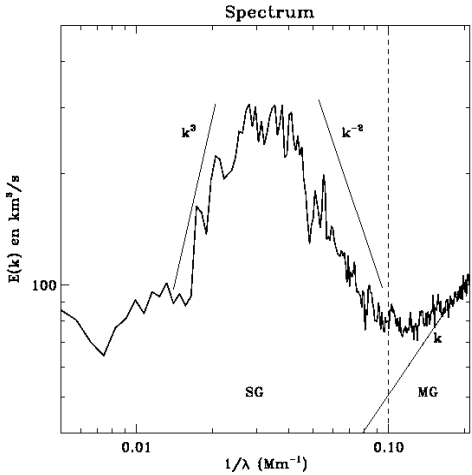,width=9 cm}}
\caption[]{  Kinetic energy spectra of the horizontal velocities. The vertical dashed line 
emphasizes the 10 Mm scale, which is usually taken as upper limit of mesogranular scale. Two power 
laws are shown on each side of the peak, as well as that of the small-scale noise.}
\label{spectra}
\end{figure}

The spectrum shown in Fig.~\ref{spectra} is very similar to 
that exhibited in Fig. 2  of \cite{RMRRBP2008}. It exhibits the kinetic energy 
contained in the supergranulation. With SDO data, we find the maximum of the spectral 
density  at a wavelength of 35.8~Mm quite close to 36~Mm found by \cite{RMRRBP2008} and 
the FWHM of the peak indicates that supergranulation occupies  the range of scales of 
[19, 56]~Mm  compared to  [20, 75]~Mm for \cite{RMRRBP2008}. Our time sequence is shorter 
by a factor 3.75 but our field of view is 3.5 larger ($400~\arcsec\times300~\arcsec$ 
for \citealt{RMRRBP2008}). 
Like Rieutord et al., we  indicate the best-fit power-law (3 and -2), which mimics the sides 
of the supergranulation spectral peak. At small scales below $k=0.1 Mm^{-1}$ we observe 
$ E(k) \propto k$ due to the decorrelated random noise. Our present result confirms 
the previous findings.

\section{Solar differential and meridional flow determination}

There is a long history of inconsistent measurements of solar rotation
\citep{Beck2000}.  One of the first applications of the CST algorithm
on HMI/SDO data is to measure the horizontal velocities over 24h on the
central meridian. Owing to computational time, we used a band of 
$50~\arcsec$ along the central meridian and 48 time-sequences of 30 min. 
Then, 98 spatial windows of $50~\arcsec\times20~\arcsec$ were used to obtain
the horizontal flow fields at different latitudes for each time-sequence of 30~min. Finally, we 
averaged all 30~min time-sequences for each latitude to obtain the final Vx and Vy
over 24h. The Vx are corrected for the Earth's orbital motion to obtain the sidereal 
rotation of the Sun, and Vy was also corrected for the B0 evolution during 24h.
This is the first time that solar rotation has been determined from granule displacement 
measurements over the full solar disk. Fig.~\ref{rotdif} 
shows the differential rotation computed in this way for four different dates close to 
the solar minimum.

 The standard deviation of the velocity field close to the disk center  is 
0.06~km s$^{-1}$ (24h sequence). Owing to the change in size 
of the Sun in diameter of (1.4 pixel on the detector) (SDO Earth orbit) and also to the B0 
evolution during 24h, the field of good measurement was reduced to  $-50\degr~ and ~+45\degr$  
in longitude and $\pm50\degr$ in latitude around the disk center. This limitation will be 
improved in the near future (new observations on 10 December 2011) by reducing the effects  
of  dilatation and contraction of the Sun's  diameter on the CCD.

Our results of April 10, 2010 agree with the spectroscopic 
determination of the solar rotation of \cite{HH70} but the rotation seems slightly  
faster  on the other dates with an equatorial velocity of 1.99 km~s$^{-1}$. 
The daily Wolf numbers are 8, 14, 33, and 26 and  8, 16.1, and 41.6 monthly. We expect differences with
\cite{HH70}, simply because we are not studying the same time period. The
solar differential rotation appears to change in time particularly at
high latitudes, but we need a more extensive analysis to confirm these variations.

\begin{figure}
\centerline{\psfig{figure=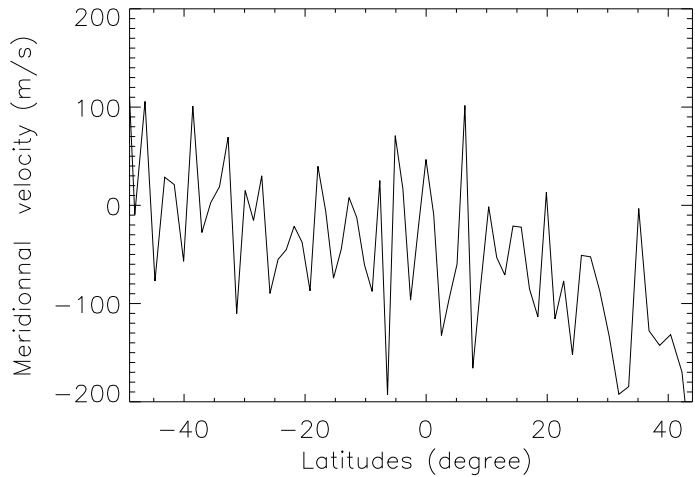,width=10 cm}}
\caption[]{ Meridionnal velocity at different latitudes across the central meridian 
on April 10, 2010, quiet Sun. }
\label{vel_meridian}
\end{figure}

 The meridional flows with Vrms = 90 m~s$^{-1}$, given by the Vy component for 24h, is plotted 
in Fig.~\ref{vel_meridian} for a quiet Sun on  April 10 2010 between $-50\degr~ and ~+45\degr$ 
in latitude. Because the Vrms of the mean meridional flow is 90 m~s$^{-1}$, the expected 
signal of 20 m~s$^{-1}$ is completely hidden. Several days are required to reduce the Vrms amplitude 
and determine the meridional velocity. This work is scheduled for the near future 
because we must also include the Sun's diameter evolution on the SDO detector over a long period 
of time (Earth orbit).

\section {Discussion and conclusions}

Determining horizontal velocity fields on the solar surface 
is crucial for understanding the dynamics of the phostosphere, the 
distribution of magnetic fields and its influence on the structures 
of the solar atmosphere (filaments, jets, etc.).

We have shown that the reconstruction of the velocity field can be achieved over the Sun's 
entire visible surface with HMI/data by using CST. From this reconstruction, by using the multi-resolution
analysis, one can also compute the velocity field at different scales along with
its derivatives such as divergence and curl. An uncertainty in the velocity of
about 0.25~km~s$^{-1}$ for a time sequence of 30~min and a mesh of 2.5
Mm is acceptable if compared to the granule velocities which range between
0.3~km~s$^{-1}$ and 1.8~km~s$^{-1}$. The kinetic energy power spectra obtained on a large field of 
the Sun confirms the previous quantitative studies of the dynamic of the Sun's surface. The first 
determination of the solar rotation using granule tracking 
demonstrates the potential of our method for studying the evolution of solar rotation on short-time scales. 
The method can also be used to remove the solar rotation for 
studying smaller scale evolution (supergranulation flows, etc.).

In a near future, it will be possible to derive the velocities on the
solar surface in the local coordinates system ($V_r, V_\theta, V_\phi$)
from the line-of-sight component velocity (Vdop) and  horizontal velocity
fields on the solar surface ($V_x$,$V_y$). In particular, it will be 
possible to compare the spherical analysis of the Sun's motions to those of 
numerical simulations, thus improving our knowledge of the physical properties 
of the Sun.

 Contemplating the CST method and the HMI/SDO data, many applications can be envisioned on
time scales ranging from 30 minutes to several days or even months with
spatial scales ranging form 2.5 Mm up to the radius of the Sun. This opens a
new field of study allowing, \textit{e.g.,} a measure of the influence of photospheric
motions on the dynamics of the outer solar atmosphere such as the triggering of filament 
eruptions, see \cite{RSMK08}. Many other applications will be possible on phenomena 
that surface motions involve.
  
Today, 12~min of processor time are required to compute 30~min of solar
observations.  Improvement of the code is in progress. Currently,
the Fortran90 CST code is parallelized in a shared memory approach
involving OpenMP. We experienced good scaling on a production data set
(full sun) with 73.16\% parallel efficiency on 24 cores. Tests were conducted
on an SGI Altix UV ccNUMA node with 96 cores and 1TB of RAM.
This CST code (\textit{Fortran90}) will be implemented in the
German Data Center for SDO and in the Joint Science Operations Center,
and will be available to solar community.

\begin{acknowledgements}

We thank the \textit{Hinode}/SOT team for assistance in acquiring and
processing the data.  \textit{Hinode} is a Japanese mission developed
and launched by ISAS/JAXA, collaborating with NAOJ as a domestic
partner, NASA and STFC (UK) as international partners. Support for
the post-launch operation is provided by JAXA and NAOJ (Japan), STFC
(U.K.), NASA, ESA, and NSC (Norway).  We thank the HMI team members for
their hard work.We thank the German Data Center for SDO and BASS2000
for providing HMI/SDO data. The German Data Center for SDO is supported 
by the German Aerospace Center (DLR). We thank F. Rincon for comments.
This work was granted access to the HPC resources of CALMIP under the
allocation 2011-[P1115].  This work was supported by the CNRS Programme
National Soleil Terre. We thank the anonymous referee for suggestions
and careful reading of the manuscript.

\end{acknowledgements}

\bibliographystyle{aa}
\bibliography{biblio}

\end{document}